\documentclass[review]{elsarticle}

\usepackage{amsmath}
\usepackage{amssymb}
\usepackage{xcolor}
\usepackage{graphicx}
\usepackage{physics}
\usepackage{booktabs}
\usepackage[version=4]{mhchem}

\usepackage{hyperref}

\newcommand{\onlinecite}[1]{\hspace{-1 ex} \nocite{#1}\citenum{#1}} 

\makeatletter
\def\ps@pprintTitle{%
   \let\@oddhead\@empty
   \let\@evenhead\@empty
   \let\@oddfoot\@empty
   \let\@evenfoot\@oddfoot
}
\makeatother
\begin{document}

\begin{frontmatter}

\title{Predictive design and experimental realization of InAs/GaAs superlattices with tailored thermal conductivity}

\author[TUW]{J. Carrete}
\author[CEA,GA]{B. Vermeersch}
\author[JKU]{L. Thumfart}
\author[Delaware]{R. R. Kakodkar}
\author[Parma] {G. Trevisi}
\author[Parma] {P. Frigeri}
\author[Parma] {L. Seravalli}
\author[Delaware]{J. P. Feser}
\author[JKU]{A. Rastelli}
\author[CEA,GA]{N. Mingo\corref{corresp}}
\ead{natalio.mingo@cea.fr}
\cortext[corresp]{Corresponding author}

\address[TUW]{Institute of Materials Chemistry, TU Wien, A-1060 Vienna, Austria}
\address[CEA]{LITEN, CEA-Grenoble, F-38054 Grenoble, France}
\address[GA]{Universit\'e Grenoble Alpes, F-38000 Grenoble, France}
\address[JKU]{Institute of Semiconductor and Solid State Physics, Johannes Kepler University Linz, A-4040 Linz, Austria}
\address[Delaware]{Department of Mechanical Engineering, University of Delaware, Newark, 19716 Delaware, USA}
\address[Parma]{IMEM - CNR Institute, Parco Area delle Scienze 37/a, Parma 43124, Italy}

\begin{abstract}
  We demonstrate an \textit{ab-initio} predictive approach to computing the thermal conductivity ($\kappa$) of InAs/GaAs superlattices (SLs) of varying period, thickness, and composition. Our new experimental results illustrate how this method can yield good agreement with experiment when realistic composition profiles are used as inputs for the theoretical model. Due to intrinsic limitations to the InAs thickness than can be grown, bulk-like SLs show limited sensitivity to the details of their composition profile, but the situation changes significantly when finite-thickness effects are considered. If In segregation could be minimized during the growth process, SLs with significantly higher $\kappa$ than that of the random alloy with the same composition would be obtained, with the potential to improve heat dissipation in InAs/GaAs-based devices.
  \end{abstract}


\end{frontmatter}


\newcommand{\dg}{\ensuremath{\;^{\mathrm{\circ}}\mathrm{C}}}

\section{Introduction}
The problem of heat management has been brought to the forefront by advances in miniaturization and the consequent increase in power dissipation density. The lifetime and performance of some devices, particularly in areas like power electronics, can be dramatically enhanced by improved heat dissipation. On the other hand, there are applications, like thermoelectric conversion, where the goal is to achieve a low thermal conductivity while maintaining a good electronic mobility. In this context, nanostructured materials such as semiconductor SLs have opened up a broad avenue of possibilities to tailor the thermal conductivity of functional materials. For example, the thermal conductivity of SiGe-based thermoelectric materials can be significantly reduced by replacing the disordered alloys with crystalline SLs of similar $\kappa$ but much lower Ge content \cite{SiGe}. Similarly, nanodot SLs have also been shown to increase the thermoelectric figure of merit of III-V semiconductors \cite{nanodot}.

A main challenge of nanostructure design is to quantify this relationship between structure and the various functional properties. In particular, for a SL it is important to determine to what extent the property of interest can be modulated by suitable choice of the structural parameters. For example the thermal conductivity can be strongly affected by the SL's compositional distribution\cite{huxtable_thermal_2002,landry_complex_2008,huberman_disruption_2013,chalopin_thermal_2012}. It is critical to take into account that segregation happens during SL growth, meaning that even if the pure components are deposited one at a time the resulting profile will not correspond to a digital alloy \cite{Haxha, Godbey,Muraki}. This can totally change the thermal conductivity of the structure as we recently demonstrated for SiGe SLs \cite{SiGe}. A crucial question for any new system is then to determine the limits of tunability by structural design. In other words, how much can $\kappa$ of a crystalline SL be lowered or increased with respect to the disordered alloy of equivalent concentration, for realistic composition profiles? Here we address this question in the case of InAs/GaAs SLs.

InGaAs-based alloys and nanostructures are of particular technological relevance due to their use in infrared photodetectors \cite{infrared}, MOSFETs and HEMTs \cite{Ajayan}. Because of the big role played by heat dissipation in the efficiency and reliability of those devices, it is important to assess whether thermal transport can be significantly optimized by changing the SL parameters, and in particular whether techniques like digital alloying can lead to higher $\kappa$ than that of a disordered alloy. Here we show that this is not the case for InAs/GaAs SLs of sufficient thickness. Our calculations of $\kappa$ for very thick superlattices with realistic composition profiles indicate that segregation during growth is a limiting factor to the variability of $\kappa$, and we present experimental measurements in good agreement with those calculated results. However, finite-size effects can exacerbate the influence of the composition profile in superlattice thin films; in particular, if segregation could be reduced, a digital alloy could display $\kappa$ values up to 25\% higher than the corresponding disordered alloy.

\section{Experimental methods}

\subsection{Growth and characterization}
\begin{figure}[htb]
 \begin{center}\leavevmode
 \includegraphics[width=8.cm]{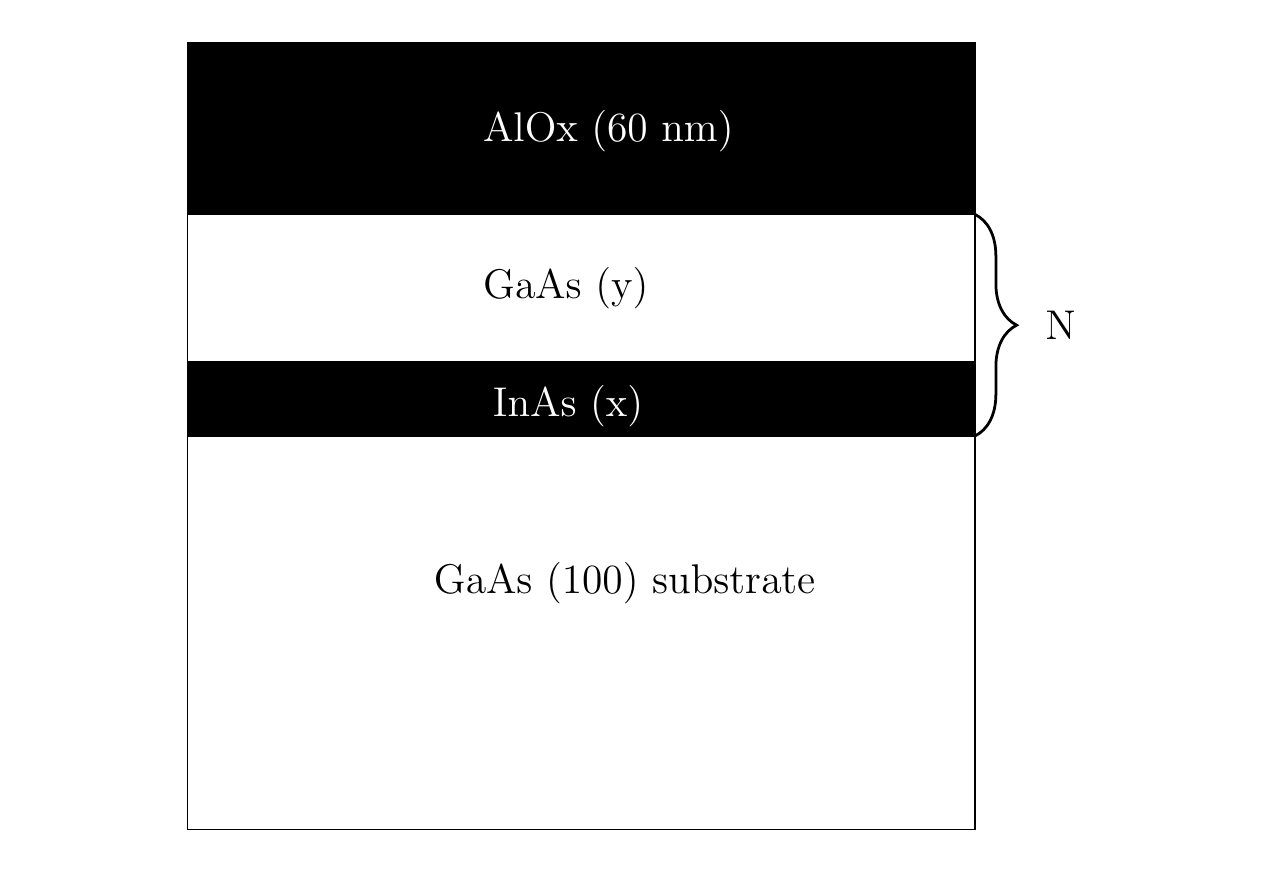}
 \caption{Schematic representation of a sample structure. The superlattice consists of N repeats of InAs/GaAs layers with nominal thickness x/y and is grown on GaAs(001) substrates via MBE. On top of the structure, a layer of AlOx is deposited via ALD to provide electrical insulation for 3$\omega$ measurements. }
 \label{fig:Struct}
\end{center}
\end{figure}

InAs/GaAs SLs are grown on GaAs(001) substrates in in two independent molecular beam epitaxy (MBE) setups, yielding fully consistent results. The samples are deoxidized and a $100\;\mathrm{nm}$ GaAs buffer is grown at $600\dg$. During the subsequent pause in growth, the temperature is reduced to $480\dg$,  which is suitable for InAs deposition. The superlattices are then deposited at this growth temperature following the pattern illustrated in Fig. \ref{fig:Struct}. To ensure uniformity across the two-inch wafer, the substrate is continuously rotated during the whole growth process.

To determine the influence of the film thickness $t_{\mathrm{film}}$ and composition of the layered structure on the thermal conductivity, the  InAs and GaAs layer thicknesses as well as the number of periods are varied individually. The parameters for all grown samples are summarized in Table \ref{tbl:samples}. The InAs content and film thickness are constrained in order to avoid the formation of additional phonon scattering centers due to dislocations, as InAs is strongly strained when grown on GaAs. The critical film thickness below which one can assume pseudomorphic growth for a given composition has been theoretically modeled by Maree et al. \cite{Maree}. 

\begin{table}[h]
  \centering
  \begin{tabular}{l r r r}\toprule
    Sample & Number of periods &InAs thickness &GaAs thickness \\
           & &x ($\mathrm{ML}$) &y ($\mathrm{nm}$)\\\midrule
    A&46& 1.3 & 10 \\
    B&46& 0.8 & 6 \\
    C&30& 1.3 & 6 \\
    D&30& 1.3 & 4.4 \\
    E&46& 1.3 (1.36)& 6 (6.09)\\
    F&70& 1.3 (1.38)& 6 (6.04)\\
    G&70& 0.8 (0.86)& 6 (5.98)\\
    H&100& 0.8 (0.87)& 6 (5.98)\\
    I&130& 0.8 (0.82)& 6 (5.99)\\\bottomrule
  \end{tabular}
  \caption{List of all grown samples with their numbers of periods and the thicknesses of each kind of layer. The thicknesses $x$ and $y$ of the InAs and GaAs layers are provided in monolayers (ML) and nm respectively. The general sample structure is sketched in Fig. 1. Values between parentheses correspond to direct measurements, and can differ slightly from the nominal values.}
  \label{tbl:samples}
\end{table}

 \subsection{Thermal characterization}

Thermal conductivities were measured using both the differential $3\omega$-method as described by Lee and Cahill \cite{CahillLee} and time-domain thermoreflectance (TDTR). In both measurement methods the thermal conductivity is obtained after heating and subsequent temperature measurement of the surface. Heat dissipates through the film and the substrate beneath. For the preparation of the $3\omega$-measurement, a $9\;\mathrm{\mu m}$-wide and $120\;\mathrm{nm}$-thick gold layer was deposited to act as both a heater and a thermometer. As the $3\omega$-method is based on an electrical measurement, leakage currents into the sample need to be avoided. Therefore $60\;\mathrm{nm}$ of \ce{AlO_x} were deposited on top of the film as an insulating layer.
A reference sample with the same kind of substrate and an identical dielectric layer on top was processed in parallel to enable extraction of the SL-film thermal conductivity, so that differences in the surface temperatures of both samples due to heating can be attributed to the different thermal conductivity of the SL with respect to the substrate.

In addition to the thermal conductivity of the individual layers, the surface temperature also depends on the heating power, the heater width, and the layer thicknesses. The SL layer thickness and the average GaAs/InAs layer thicknesses are determined through x-ray diffraction (XRD). The heater width is measured by atomic force microscopy (AFM).

For the  thermal resistance measurements, the samples were placed in vacuum to minimize thermal losses due to convection. The surface was heated with $15\;\mathrm{W\,m^{-1}}$ of power linear density, with heating frequencies ranging from $40\,\mathrm{Hz}$ to $4\;\mathrm{kHz}$, and at room temperature.

In the $3\omega$-method the heat transport is effectively two-dimensional, with components parallel and perpendicular to the surface. For large heater widths ($b$) compared to the SL film thickness (t$_\mathrm{film}$), $b\gg t_{\mathrm{film}}$, and SL thermal conductivities ($\kappa_\mathrm{film}$) substantially lower than substrate thermal conductivity ($\kappa_\mathrm{substrate}$), $\kappa_{\mathrm{film}}\ll\kappa_{\mathrm{substrate}}$, in-plane transport in the SL is negligible. However, for our InAs/GaAs SLs, where the thermal conductivities of the substrate and the film are comparable and some films have thicknesses of about $1\;\mathrm{\mu m}$, in-plane transport has to be taken into account. The surface temperature can be fitted taking the  heater width, the substrate temperature, and the film thickness as input parameters. This is done by solving the 2D heat transfer equation with the thermal conductivities of the film and the substrate as adjustable parameters. A matrix formalism is used for both regions (film and substrate) with the boundary conditions of temperature continuity and energy conservation.

Thermal measurements between $30$ and $700\;\mathrm{K}$ were also performed using a two-tint implementation of time-domain thermoreflectance (TDTR) \cite{KangAndCahill} for a selection of samples (A, B, C and D in Table \ref{tbl:samples}), as a way to independently check the results of the measurements and post-processing. TDTR is a pump-probe technique where a pump beam is used to heat the sample and a probe beam measures the surface temperature as a function of time.  Experiments were conducted using a mode-locked Ti:sapphire laser with $\lambda =785\;\mathrm{nm}$ and a repetition rate of $76\;\mathrm{MHz}$. The pump beam was modulated at a frequency of $12.6\;\mathrm{MHz}$, and with a focused beam with $1/e^2$ spot size of $5\;\mathrm{\mu m}$. We used a total laser power of around $90\;\mathrm{mW}$, which we estimate gives a maximum steady-state temperature rise $<0.5\%$ of the absolute temperature. An aluminum layer of thickness $70\;\mathrm{nm}$ was coated on the sample to act as a thermal transducer. The measured change in reflectivity as function of time delay was fitted to an analytically obtainable thermal model. Data reduction \cite{Cahill_DataReduction} is performed between $300$ and $4000\;\mathrm{ps}$, treating the thermal conductivity of the superlattice layer and the thermal conductance of the Al/superlattice interface as the only fitting parameters. The data reduction requires prior knowledge of the heat capacities of GaAs and InAs, which were obtained from the literature \cite{Touloukian}.  Error bars for the measurements are estimated by refitting the data introducing uncertainties in the known parameters used as inputs to the thermal model, with typical error bars found to be $\sim 13\%$.

The TDTR and $3\omega$ methods were found to agree within their respective uncertainties. In the remainder of this article, TDTR data is used if available (samples A, B, C and D), and $3\omega$ data is used otherwise.

\section{Theory}

\subsection{Ab-initio thermal conductivity calculations}

The lattice thermal conductivity of single crystals and bulk alloys can be accurately predicted \textit{ab initio} through the numerical solution of the linearized phonon Boltzmann transport equation (BTE) \cite{BroidoAPL}. The general approach has been described, for example, in Ref.~\cite{sheng}. In brief, the procedure requires the ab-initio calculation of the second- and third-order interatomic force constants for the material at hand. A dense regular grid is then constructed in order to sample the Brillouin Zone of the crystal. From those constants, the phonon frequencies $\omega_{\alpha}\left(\mathbf{q}\right)$ and eigenvectors $\ket{\mathbf{q}, \alpha}$ are computed for each point on the grid ($\mathbf{q}$) and each phonon branch ($\alpha$). As the final ingredient, the scattering amplitudes for each phonon mode are evaluated from knowledge of all different types of scattering, taking into account the atomic structure of the system. Those include two-phonon processes with amplitudes $\Gamma_{\lambda\lambda'}$, enabled by any breakdown of perfect periodicity such as crystallographic defects, and three-phonon emission (adsorption) processes with amplitudes $\Gamma^{- (+)}_{\lambda\lambda'\lambda''}$ derived from any deviation of the Hamiltonian from perfect harmonicity. In the symbols for these amplitudes, a single index $\lambda$ stands for both a $\mathbf{q}$ point and a branch index $\alpha$. We keep this notation for the remainder of this article.

For pure single crystals, scattering sources are three-phonon anharmonic processes \cite{threeph}, and two-phonon isotope scattering \cite{tamura_isotope_1983}. The same approach introduced to deal with mixtures of isotopes can also be extended to alloys. This seemingly drastic extrapolation is at the root of the virtual crystal approximation (VC) which has in fact proved to be rather effective for many alloys \cite{VCA}. More complex defects like vacancies\cite{katcho_effect_2014}, substitutions \cite{katre_unraveling_2016}, or dislocations \cite{wang_ab_2017} are yet another source of scattering that can be characterized \textit{ab-initio} from the atomic level structure of the defect using Green's functions. With all these elements, the linearized BTE can be solved to yield the non-equilibrium phonon distribution in the material subject to a small temperature gradient. Once the distribution is known, transport properties like the thermal conductivity tensor are easily computed \cite{sheng}.

Phonon scattering in SLs has very specific features that set this class of systems apart from the ones described above. Those features stem from the presence of a privileged growth direction, which creates a complex interplay between order and disorder. To state this in more precise terms, we need to introduce a site-by-site description of an SL as a three-dimensional structure. In the spirit of the virtual crystal approximation, we neglect the structural distortions around particular sites, and model the SL as a structurally periodic medium, where each atom is placed at the crystalline sites of a hypothetical zinc-blende crystal with an effective lattice parameter computed as the weighted average of the two lattice parameters of the pure compounds: $a = x_{\ce{InAs}}a_{\ce{InAs}} + x_{\ce{GaAs}}a_{\ce{GaAs}}$. Each primitive unit cell in the virtual crystal contains a $\ce{\mathcal{X}As}$ pair, where $\mathcal{X}\in\left\lbrace \ce{In}, \ce{Ga} \right\rbrace$. We can characterize the directionality of a SL by a reciprocal-space vector or, equivalently, by the Miller indices $(hkl)$ representing the family of planes that make it up. It is convenient to base our system of coordinates on this concept: Let $i$ be an integer coordinate selecting a particular plane from the family, and let $j$ and $k$ be integer coordinates selecting unit cells in the layer; we introduce an indicator variable $X\left(i, j ,k\right)$ that takes the value $1$ if $\mathcal{X} = \ce{Ga}$, and $0$ if $\mathcal{X} = \ce{In}$, and the local decomposition

\begin{equation}
  X\left(i, j ,k\right) =: \bar{X}\left(i\right) + \delta X \left(i, j\right).\label{eqn:decomposition}
\end{equation}

Here, $\bar{X}\left(i\right)$ is the average of $X\left(i, j ,k\right)$ over layer $i$, and hence $\delta X \left(i, j\right)$ is a local in-plane deviation. Obviously, $\bar{X}$ has the same periodicity as the superlattice along its growth direction. This term varies along $z$ in lengths of a few nm, and it can scatter long wavelength phonons. Moreover, since it does not depend on the in-plane directions, scattering related to this term will not change the parallel momenta of the phonons. The second term, $\delta X(i,j,k)$, varies quickly in all three directions, in scales comparable to the interatomic distance, so it preferentially scatters short wave phonons, and it is not constrained by parallel momentum conservation.

To quantitatively evaluate the phonon scattering rates induced by these terms, we consider the perturbation that they produce on a virtual crystal with atomic masses and force constants corresponding to the concentration-weighted averages of those from InAs and from GaAs, defined just like the effective lattice constant above. The force constants of InAs and GaAs are rather similar, and differences in their phonon dispersions stem primarily from the different masses of their atoms. Therefore, the largest contribution to the perturbation in the phonon Hamiltonian comes from mass differences, leading to a frequency-dependent diagonal perturbation on the In/Ga sites of the form $\left(V_{l,m}\right)^{\alpha,\beta} = \delta_{l,m}\delta_{\alpha,\beta}V_l $, where $l$ and $m$ label the atoms, $\alpha$ and $\beta$ label Cartesian directions, and  \cite{Kundu}

\begin{equation}
  V_l = \frac{M_l-M^{\mathrm{VC}}_l}{M^{\mathrm{VC}}_l}\omega^2.
\end{equation}

\noindent If $X_l$ is the value of $X \left(i, j, k\right)$ in the unit cell where $l$ belongs, then $M_l$ can be expressed simply as  $M_l=m_{\ce{Ga}}X_l+m_{\ce{In}}\left(1-X_l\right)$. Combining this with the decomposition in Eq. \eqref{eqn:decomposition}, we can identify two contributions to the perturbation, $V_l = V_l^a + V_l^b$, with each term defined as:

\begin{subequations}\label{eqn:contributions}
  \begin{align}
    V_l^b &= \frac{\bar{M}_l - M^{\mathrm{VC}}_l}{M^{\mathrm{VC}}_l}\omega^2,\text{ with }\bar{M}_l=m_{\ce{Ga}}\bar{X}_l+m_{\ce{In}}\left(1-\bar{X}_l\right),\label{eqn:contributions:b}\\
    V_l^a &= \frac{\delta M_l}{M^{\mathrm{VC}}_l}\omega^2    ,\text{ with }\delta M_l=\left(m_{\ce{Ga}}-m_{\ce{In}}\right)\delta X_l.\label{eqn:contributions:a}
\end{align}
\end{subequations}

\noindent The ``barrier'' term $V_l^b$ and the ``alloy'' term $V_l^a$  are very different in their nature. $V_l^b$ changes only along the growth direction, and represents the effect of the average mass profile scattering due to nanostructuring. $V_l^a$ represents the scattering due to disordered fluctuations in mass at each atomic site.

The two-phonon scattering amplitudes due to this perturbation can be calculated in the framework of converged time-independent perturbation theory. Specifically, the required ingredient is the causal t matrix $\mathbf{t}^+$,  since $\Gamma_{\lambda\lambda'}\propto \left\vert \mel{\lambda'}{\mathbf{t}^+}{\lambda}\right\vert^2$. The total contribution of the perturbation to the scattering rate of mode $\lambda$ can be extracted from a single matrix element thanks to the optical theorem:

\begin{equation}
  \tau^{-1}_{\lambda, \mathrm{SL}} = -\frac{1}{\omega_\lambda} \Im \mel{\lambda}{\mathbf{t}^+ }{\lambda}
  \label{eqn:optical}
\end{equation}

\noindent The causal t matrix is connected to the perturbation through a formal identity involving the causal Green's function ($\mathbf{g}^+$) of the unperturbed system:

\begin{equation}
  \mathbf{t}^+=\mathbf{V} +\mathbf{V} \mathbf{g}^+ \mathbf{V} +\mathbf{V} \mathbf{g}^+ \mathbf{V} \mathbf{g}^+ \mathbf{V} +\ldots
  \label{eqn:geometric}
\end{equation}

\noindent Since $V^a_l$ varies randomly in space, as a first approximation we can neglect its correlations to disregard any cross terms involving $V^a_lV^a_m$ or $V^b_lV^a_m$ for $l\ne m$. This allows us to approximate the t matrix as the sum of separate contributions from barriers and alloy disorder, as $\mathbf{t}^+ \simeq \mathbf{t}^{+b} + \sum_l \mathbf{t}^{+a}_{l,l}$. When inserted into the optical theorem \eqref{eqn:optical}, this also yields a sum of separate alloy and barrier contributions to the total scattering rates: $\tau_{\lambda, \mathrm{SL}}^{-1}= \tau^{-1}_{\lambda,a}+\tau^{-1}_{\lambda,b}$. Thus the problem of phonon scattering in superlattices is reduced to two separate problems: scattering due to isolated point defects, and scattering due to an average mass distribution modulated only along one direction. An additional source of scattering comes from the two external interfaces of the finite-length SL. This effect is difficult to model without a detailed knowledge of the character of the interfaces, which is typically not available. However, as we show in the next section, assuming perfectly diffuse interfaces yields good results.

Experience in dealing with alloys has shown that the effect of mass defect scattering can be well described in the Born approximation, which amounts to taking $\mathbf{t}^t \simeq \mathbf{V}$ in each of the matrix elements $\left\vert \mel{\lambda'}{\mathbf{t}^+}{\lambda}\right\vert^2$. This leads to a formula analogous to the one developed by Tamura for phonon scattering due to isotopic mass disorder in single crystals:

\begin{equation}
  \tau^{-1}_{\lambda,a,l} = \frac{\pi \omega^2_{\lambda}}{2} \left(\frac{\delta M_l}{M_l^{\mathrm{VC}}}\right)^2\sum\limits_{\lambda'}\left\vert \braket{\lambda'}{l} \braket{l}{\lambda}\right\vert^2\delta\left(\omega_{\lambda'} - \omega_{\lambda}\right),
  \label{eqn:tamura}
\end{equation}

\noindent where $\braket{l}{\lambda}$ denotes the projection of the vibrational eigenvector $\ket{\lambda}$ on the degrees of freedom of the In/Ga atom of the virtual crystal, and the sum over $\lambda'$ includes both a sum over branches and an average over the Brillouin zone of the virtual crystal. The only factor in Eq. \eqref{eqn:tamura} that depends on the perturbation (and hence on the site) is the square of the local deviation from the average layer mass. Thus, to obtain the final effective contribution to scattering from the ``alloy'' part of the perturbation we only need to compute the expected value of that term over all sites and all configurations of the alloy. As can be readily checked,

\begin{equation}
  \begin{aligned}
    \mathbf{E}\left\lbrace \tau_{\lambda, a}^{-1}\right\rbrace &\propto \mathbf{E}\left\lbrace \left[\delta M \left(i, j, k\right)\right]^2 \right\rbrace = \frac{1}{N_{\mathrm{period}}} \sum_{i=1}^{N_{\mathrm{period}}} \left\lbrace \overline{M^2}\left(i\right) - \bar{M} ^2\left(i\right) \right\rbrace=\\
    &=\frac{1}{N_{\mathrm{period}}} \sum_{i=1}^{N_{\mathrm{period}}}\left\lbrace \bar{X}\left(i\right) \left[m_{\ce{Ga}} - \bar{M}\left(i\right)\right]^2 + \left[1- \bar{X}\left(i\right)\right] \left[m_{\ce{In}} - \bar{M}\left(i\right)\right]^2 \right\rbrace =\\
    &=\frac{\left(m_{\ce{Ga}} - m_{\ce{In}}\right)^2}{N_{\mathrm{period}}} \sum_{i=1}^{N_{\mathrm{period}}}\left\lbrace \bar{X}\left(i\right) \left[1 - \bar{X}\left(i\right)\right]\right\rbrace
  \end{aligned}
  \label{eqn:variance}
\end{equation}

\noindent In other words, alloy scattering is proportional to the mass variance of a layer, averaged over all layers in a period of the SL. It is more intense when the concentrations of InAs and GaAs in the layers are similar, and tends to zero for digital profiles, i.e., when each layer is either pure InAs or pure GaAs.

The barrier contribution to scattering requires a full Green's function treatment, because the Born approximation runs into problems as soon as the assumption $\left\lVert \mathbf{g}^+  \mathbf{V} \right\rVert \ll 1$ becomes questionable [see Eq. \eqref{eqn:geometric}]. This can happen even for localized defects when the perturbation they introduce is strong, as in some cases of substitutional impurities \cite{katre_exceptionally_2017}. However, since the matrix norm is extensive on the size of the matrix, it is also related to the spatial extent of the perturbation. In particular, the multiple interference effects that can appear when the size of the perturbation is in the same order of magnitude as the wavelength of the scattered phonons can only be captured by the full t matrix.

In the case of barrier scattering, the scatterer is translationally invariant in the $x$ and $y$ directions. Therefore, phonons scattered by $V^b$ preserve their parallel wave-vector: an incident state can only be scattered to other states with the same $\mathbf{q}_\parallel$. Thus we compute the t matrix in a hybrid real-reciprocal space representation: the direction perpendicular to the SL layers is treated in real space as a function of the layer indexes $i$ and $i'$, whereas the translationally invariant directions parallel to the layers are considered in reciprocal space, as a function of the vector $\mathbf{q}_\parallel$. The t matrix $\mathbf{t}^+$ is computed from the causal Green's function of the unperturbed virtual crystal using the synthetic form of the sum in Eq. \eqref{eqn:geometric}, i.e., $\mathbf{t}^+  = \left[\mathbf{1} -\mathbf{V} \mathbf{g}^+ \right]^{-1}\mathbf{V}$. The causal Green's function is computed in the hybrid basis as:

\begin{align}
  &g^+_{\left(i, \alpha\right), \left(j, \beta\right)} \left(\mathbf{q}_{\parallel}, \omega^2\right) =\nonumber\\ &\lim\limits_{\epsilon\rightarrow 0^+} \sum\limits_b\frac{1}{\mathrm{length}\left[L\left(q_{\perp}\right)\right]}\int\limits_{L\left(q_{\perp}\right)}                                                                                                                        \frac{\braket{i,\alpha}{q_{\perp}, \mathbf{q}_{\parallel},b}\braket{q_{\perp}, \mathbf{q}_{\parallel},b}{j,\beta}}{\omega^2 - \omega^2_b\left(q_{\perp}, \mathbf{q}_{\parallel}\right)+i\epsilon} d q_{\perp}.\label{eqn:1dgf}
\end{align}

\noindent $\alpha$ and $\beta$ are phonon branch indices, and $q_{\perp}$ is the non-conserved component of the phonon momentum. The integral runs over the straight segment in reciprocal space defined by the point $\mathbf{q}$, the growth direction, and the boundaries of the reciprocal unit cell.

Once $\mathbf{t}^+$ is known, the barrier contributions $\tau_{\lambda, b}^{-1}$ to the total scattering rates for each phonon in the 3D grid are calculated using the optical theorem \eqref{eqn:optical} and included in the solution of the Boltzmann transport equation in the relaxation-time approximation.

A potential ambiguity in the model concerns how many periods of the SL should be taken into account when building the mass perturbation representing the barriers. Physically speaking, this is connected to the distance over which phonons can be expected to maintain coherence\cite{tian_greens_2014}. Previous results for Si/Ge SLs suggest that each barrier scatters phonons independently, and coherence between barriers can be neglected. This is likely the result of small random variations in period length, and other similar random imperfections. In the current case of InAs/GaAs SLs, we have checked that the results change little whether one considers  $\mathbf{V}^b$ as the perturbation of a single period, or over two or three consecutive coherent periods, therefore justifying the use of an incoherent barrier picture.

The formalism described in this section is implemented in the almaBTE \cite{ALMA-web} open-source software package. Details of the implementation are given in Ref. \cite{ALMA}, in particular those related to the very important question of how to evaluate the integral in Eq. \eqref{eqn:1dgf} numerically. We divide each integration segment in $501$ parts to obtain fully converged results. For integrations in the 3D Brillouin zone we use $24\times 24 \times 24$ regular q-point grids.

All of the interatomic force constants required for this process are obtained from forces on atoms in particular configurations. To calculate those forces we use the plane-wave-based density-functional-theory (DFT) software VASP \cite{vasp_general_1, vasp_general_2, vasp_general_3, vasp_general_4} with projector-augmented-wave datasets \cite{vasp_paw_1, vasp_paw_2} and the Perdew-Burke-Ernzerhof approximation to exchange and correlation \cite{vasp_pbe_1, vasp_pbe_2}. We choose high energy cutoffs of $311$ and $367\;\mathrm{eV}$ for InAs and GaAs, respectively, and include their $d$ electrons in the valence. We use $5\times 5\times 5$ supercells of the primitive zincblende unit cell for all second- and third-order calculations. We employ phonopy \cite{phonopy} and thirdorder.py \cite{sheng}, respectively, to generate the displaced supercell configurations used in those calculations.

\subsection{SL growth models}

The barrier perturbation Hamiltonian requires knowledge of the average In distribution along $z$ within the SL. This inevitably depends on the nature of the growth process and on tunable parameters like growth temperature. Significant research effort has historically been devoted to the development of growth models that can provide an estimate of the SL profile based on those parameters and on the input compositions; a good review of the physical motivations and quality of the results from some of the main contenders in this area can be found in Ref.~\cite{Haxha}. In this article we refer to three classes of superlattice profiles, which we briefly describe in the following paragraphs.

The simplest of our models comprises digital SL profiles. It assumes that only pure InAs and pure GaAs are used during the growth process. If the number of In monolayers ($n_{\mathrm{ML}}$) deposited per period is an integer, the results are essentially digital alloys with a 1D structure, in which each layer contains either pure InAs or pure GaAs. For a non-integer number of In monolayers,  we build a profile with $\left\lfloor n_{\mathrm{ML}}\right\rfloor$ monolayers of pure InAs followed by a single monolayer with a GaAs concentration of $1 + \left\lfloor n_{\mathrm{ML}}\right\rfloor - n_{\mathrm{ML}}$. Digital profiles can be good models of real SLs for combinations of compounds for which interdiffusion is very limited, such as GaAs + AlAs. On the other hand, they are very far removed from realistic InAs/GaAs profiles; we use them mainly for the sake of comparison with an extreme case of what could happen if segregation were minimized.

The next model in order of complexity is based on a parameterized approximation to ion exchange during the growth process, leading to a profile of the form:

\begin{equation}
  \bar{X}\left(i\right)=\begin{cases}
    1, & \text{for } i < 1\\
    1 - \phi\left(1 - R^i\right), & \text{for } 1 \le i < n_{\mathrm{ML}} \\
    1 - \phi\left(1 - R^{n_{\mathrm{ML}}}\right)R^{i-n_{\mathrm{ML}}}, & \text{for } i \ge n_{\mathrm{ML}} \\
  \end{cases}
  \label{eqn:muraki}
\end{equation}

\noindent Here, $\phi$ is the input In content, and $R$ is the only adjustable parameter. It can be connected to an observable segregation length $\lambda$ through the formula $R = e^{-a/\left(2\lambda\right)}$, and in practice contains the effect of temperature and any other relevant parameters. We use $R=0.8$ for our growth temperature \cite{Muraki}, and refer to the results of this model as Muraki profiles, after its creator.

Finally, we also employ a kinetic model that tries to describe the evolution of concentrations in a small number of layers close to the surface through time-dependent differential equations. In particular, we use the three-layer model proposed by Godbey and Ancona. For the sake of brevity we do not reproduce the details of the model here; the parameters we used are those of Ref.~\cite{Haxha}.  By construction, this model only works when $n_{\mathrm{ML}}$ is an integer. In any other case we perform a linear interpolation between the SL profiles with $\left\lfloor n_{\mathrm{ML}}\right\rfloor$ and $\left\lceil n_{\mathrm{ML}}\right\rceil$ In monolayers.

\section{Results and discussion}
There are three main structural parameters that define the SL: the SL period length $d=x+y$, the total SL thickness $L$, and the amount of In contained in one period, $n_{\mathrm{ML}}$, given as a number of monolayers (i.e. the number of monolayers that would be filled in each period if In did not segregate). 

\begin{figure}[htb]
 \begin{center}\leavevmode
 \includegraphics[width=\columnwidth]{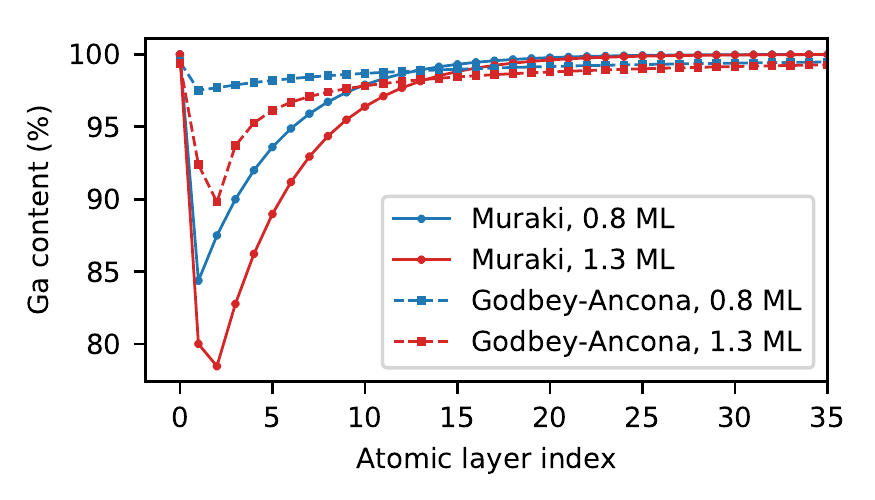}
 \caption{Composition profiles for a single $10$-nm-thick period of a InAs/GaAs SL grown using 0.8 or 1.3 monolayers of InAs, according to the two growth models employed in this paper.}
 \label{fig:profiles}
\end{center}
 \end{figure}

The number of InAs layers that can be grown in a planar InAs/GaAs SL is limited to less than $2$, the point at which 3D growth starts taking place. We have thus restricted our calculations to thicknesses in this range, in order to compare with our grown samples with $n_{\mathrm{ML}}=0.8$ and $n_{\mathrm{ML}}=1.3$ (see Table 1). The Ga content profiles calculated with the Muraki and Godbey-Ancona models are shown in Fig.~\ref{fig:profiles}. Both models predict a significant degree of segregation, with profiles that do not resemble a digital SL. This is especially marked in the case of In segregating into the newly formed Ga layers above the minimum of the Ga concentration profile. The agreement between the Muraki and Godbey-Ancona models cannot be described as quantitative, though, with the former predicting a much larger peak concentration of In.

\begin{figure}[h]
 \begin{center}\leavevmode
 \includegraphics[width=\textwidth]{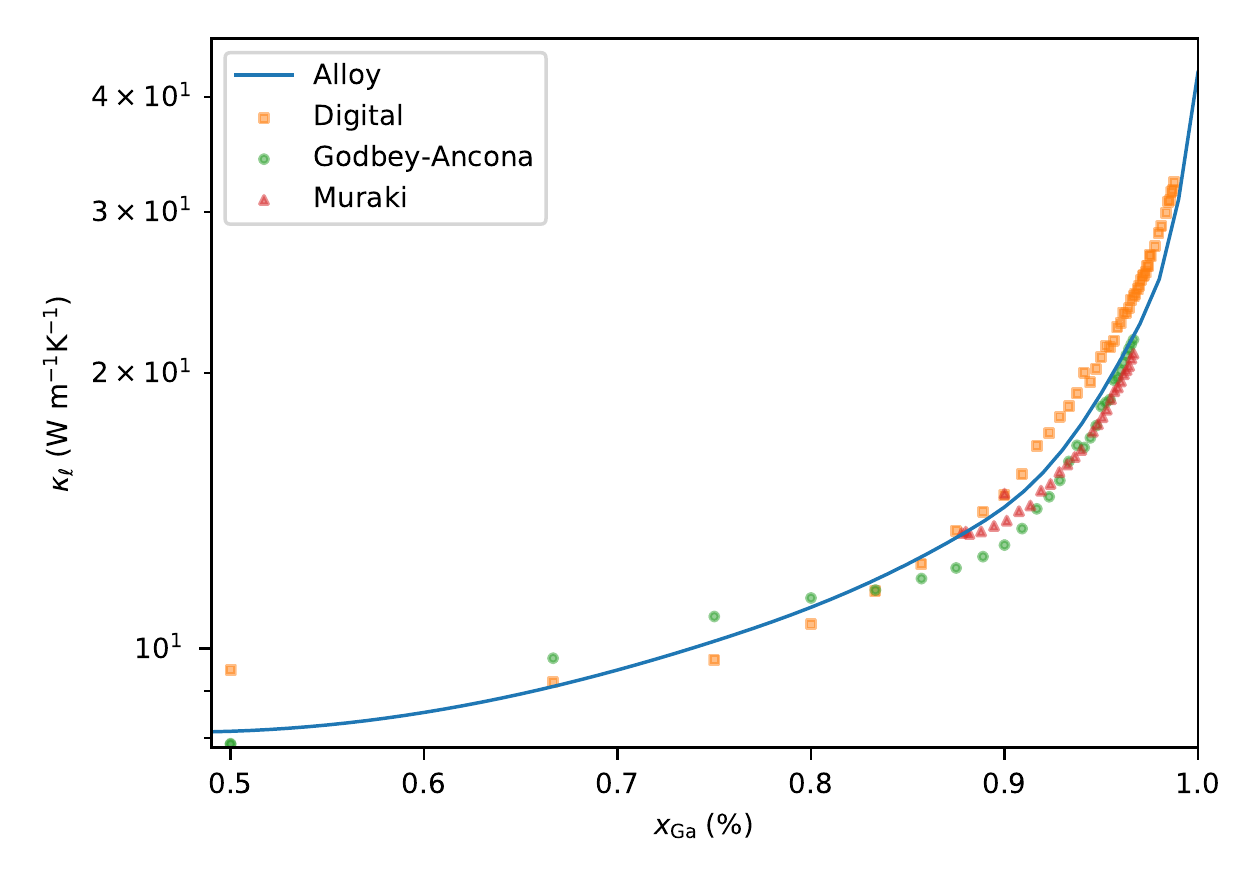}
 \caption{Calculated thermal conductivity as a function of concentration for different kinds of SL profiles and for the disordered alloy. All the profiles used for this figure contain 1 monolayer of InAs.}
 \label{fig:conductivity}
\end{center}
 \end{figure}

 As a way to assess the effect of those differences on the thermal conductivity, and more generally the effect of the composition profile on thermal transport, we calculated the value of $\kappa$ for a large set of SLs with different concentrations built using the three models discussed in the previous sections. In all cases the number of In monolayers is $n_{\mathrm{ML}}=1$, and the average composition is controlled by changing the number of Ga monolayers, which affects the period of the superlattice. All results are shown in Fig.~\ref{fig:conductivity}, for bulk-like  SLs (in the limit $L\rightarrow\infty$). The results for the completely disordered alloy throughout the range of compositions, computed using the same \textit{ab-initio} methods, are also provided for comparison.
 
 As far as bulk-like systems are concerned, Fig. \ref{fig:conductivity}  shows that the conductivity of SLs closely tracks that of the completely disordered alloy. This is due to the fact that, regardless of the concentration, there is only one InAs monolayer per period in the SL profiles presented in that figure, reflecting the experimental constraints on the number of In monolayers that can be deposited. In this regard, the (hypothetical) digital SLs with concentrations close to $50\%$ emerge as the most interesting with a view to practical applications, since they are predicted to yield the biggest deviations in $\kappa$ with respect to the disordered alloy of the same composition. This can be traced to their very short periods, insufficient to effectively scatter those wavelengths that are more relevant for heat transport. As we progress towards the Ga-rich right side of Fig. \ref{fig:conductivity}, periods become longer and barrier scattering grows in importance, causing the SL thermal conductivity to become lower than that of the corresponding alloy. For digital superlattices, however, this happens only for a brief interval of concentrations: even more Ga-rich digital SLs ($x_{\mathrm{Ga}} \gtrsim 85\%$) again have higher thermal conductivities than the alloy, since they contain long GaAs-only regions in their profiles that contribute neither to barrier scattering nor to alloy scattering. Superlattices including segregation, in contrast, have a more significant alloy contribution to scattering even for rather long periods, since none of their layers contain pure GaAs or pure InAs. Together with barrier scattering, that contribution pushes their thermal conductivity below that of the alloy for most of the points in Fig. \ref{fig:conductivity}. Interestingly, at intermediate concentrations the conductivity of SLs built using the Godbey-Ancona model is higher than the corresponding alloy, making their behavior almost completely reciprocal to that of digital SLs. In that small range of concentrations, the increased alloy scattering is not enough to compensate for the loss of barrier scattering due to diffusion.
 
Calculations and measurements on Si-rich SiGe based SLs have shown much stronger reductions of $\kappa$ below the alloy limit \cite{SiGe}. Even for 1 monolayer of Ge, the reduction reported below the alloy was about $50\%$. The Ge segregation profiles were however quite similar to those obtained for In here. The main difference is that InGaAs is a binary compound that shares a common As sublattice throughout the structure. Thus, the average mass difference distribution in the barriers is much lower than what could be expected if occupations on both sites could have been changed. Since mass disorder scattering goes with the square of the mass difference, the same barrier profile in the binary compound scatters roughly four times more weakly than in the elemental compound.

Nevertheless, the total thermal conductivity of the bulk SL is just the shallowest level of analysis of lattice thermal transport. The fact that two profiles yield the same $\kappa$ in the $L\rightarrow\infty$ does not mean that heat is being carried predominantly by similar classes of phonons. In fact, given the conceptual contrast between alloy and boundary scattering discussed in the previous sections, large differences are to be expected as a function of the degree of segregation. This has direct implications for the thermal conductivity when boundary scattering is introduced in the picture, i.e., in the frequent case where $L$ is shorter than the relevant mean free paths of phonons involved in heat transport. In a previous paper \cite{apl-thinfilms}, crystals containing significant sources of elastic phonon scattering, as those under study here, were found to display a quasiballistic regime with fractional length dependence $\kappa(L) \sim L^{\alpha}$ where $1 < \alpha < 2$ is the characteristic superdiffusion (L\'evy) exponent of the material. Ideal Rayleigh scattering would yield $\alpha = 1.75$; first-principles results for alloys \cite{apl-thinfilms} point to lower exponents ranging from $\alpha \simeq 1.66$ (\ce{\{Si/Ge\}}) to $\alpha \simeq 1.72$ (\ce{\{Al/Ga\}N}). For SL thin films it is reasonable to expect $\alpha$ to depend on the profile type due to the different scattering contributions, and thus for larger differences in thermal conductivity to appear at finite $L$.

 \begin{figure}[p!]
 \begin{center}\leavevmode
 \includegraphics[width=9 cm]{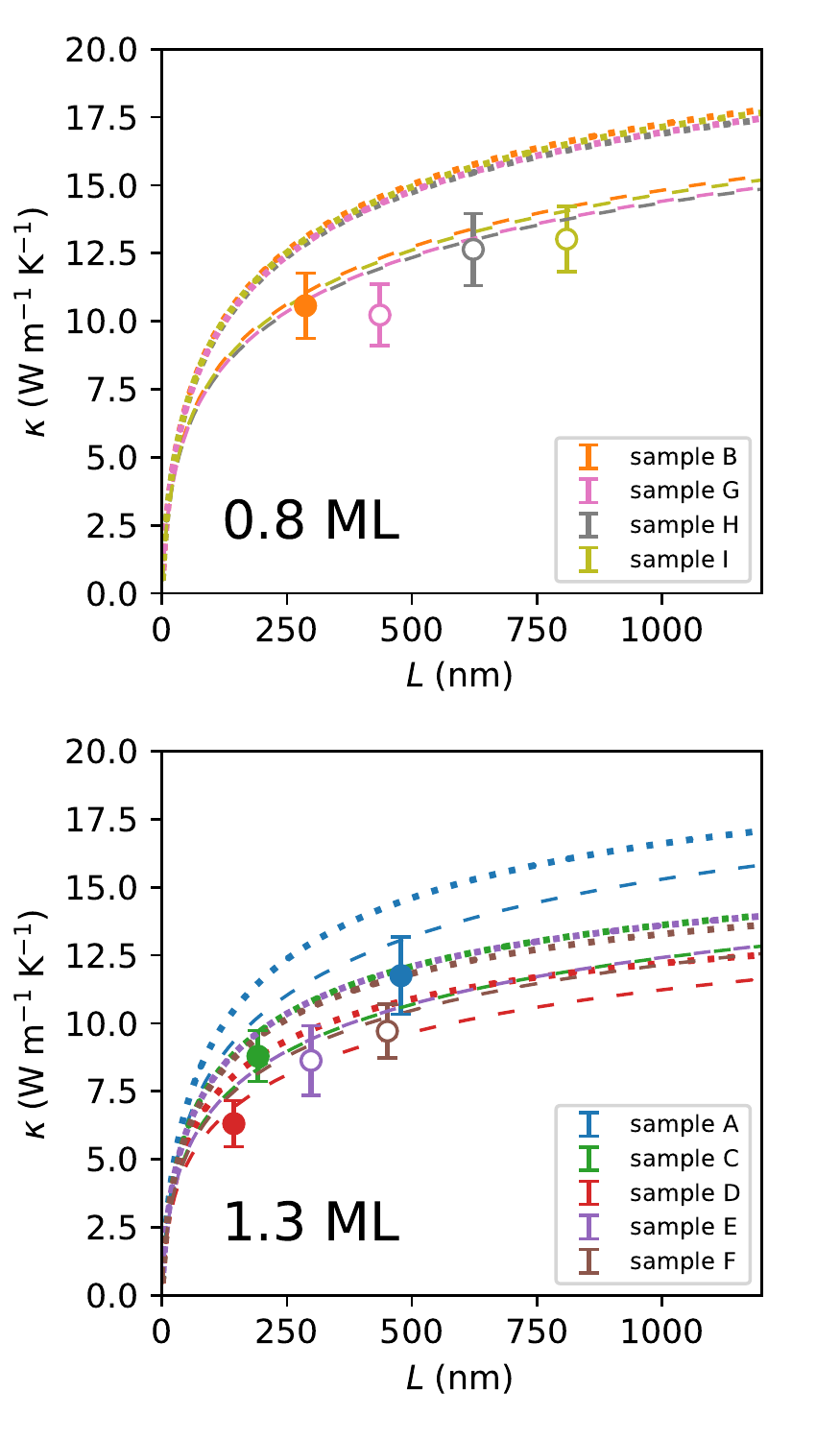}
 \caption{Comparison of experimental measurements with theoretical results when SLs are modeled using digital (dotted lines) and Muraki (dashed lines) profiles, and dependence of theoretical results on film thickness. Solid symbols correspond to TDTR measurements, and empty symbols to $3\omega$ ones.}
 \label{fig:xplane}
\end{center}
\end{figure}

\begin{figure}[h]
  \begin{center}\leavevmode
    \includegraphics[width=9 cm]{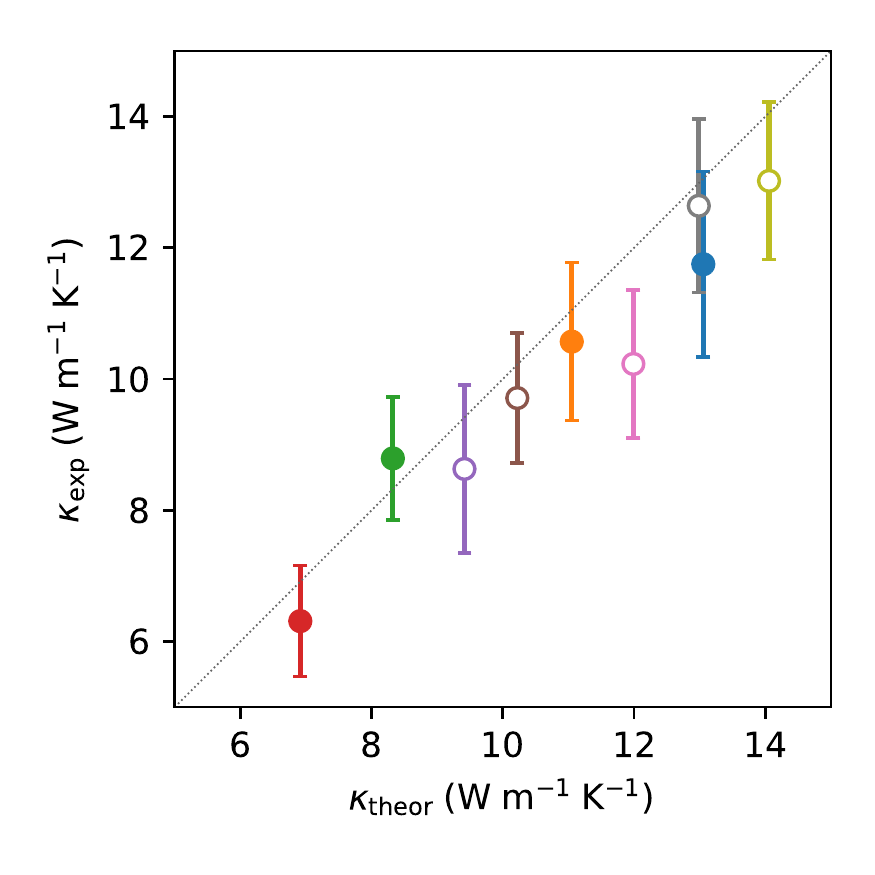}
    \caption{Comparison between theoretical predictions (using Muraki profiles) and experimental data. The colors and symbols match those in Fig. \ref{fig:xplane}.}
    \label{fig:comparison}
  \end{center}
\end{figure}

The two main panels of Fig. \ref{fig:xplane} show our experimental measurements as points in the $\left(\kappa, L\right)$ plane, compared with the theoretical $\kappa\left(L\right)$ curves computed for those SLs assuming digital profiles (left) and Muraki profiles(right). The difference in behavior is apparent, as is the fact that experimental results can only be reproduced if segregation is taken into account. Fig. \ref{fig:comparison} contains a more direct comparison between theory and experiment, showing fairly good agreement when the experimental uncertainty is taken into account, despite the simplicity of the Muraki model of SL growth and the incomplete information about the behavior of boundaries.

According to these calculations, superlattices without segregation tend to have much higher thermal conductivities in the $100 - 1000\;\mathrm{nm}$ range of thicknesses. This is connected to the fact that alloy scattering and barrier scattering tend to affect different kinds of phonons. Alloy scattering corresponds to the effect of a distribution of point-like scattering centers whose scattering cross-sections decrease rapidly with increasing wave numbers. In contrast, the concentration profile causing barrier scattering acts as an extended defect, able to efficiently scatter phonons in the $1-10\;\mathrm{nm}$ wavelength range and even below, but less effective at higher frequencies. In the vanishing-thickness limit of thin films, the mean free path of all phonons is limited by $L$. As the thickness is increased, lower-frequency phonons with long intrinsic mean free paths are still strongly affected by the boundary, while higher-frequency ones behave more like in the bulk. Since digital SL profiles provide no other scattering mechanism to depress the contribution to $\kappa$ from those higher frequencies, $\kappa\left(L\right)$ increases more abruptly.

To put this mental image in more precise terms, we can fit each of the $\kappa$ vs. $L$ to an expression of the form described in Ref. \onlinecite{apl-thinfilms}. The digital profiles in Fig. \ref{fig:xplane} correspond to values of $\alpha = 1.653\pm0.053$ (average and standard deviation of the samples) while the Muraki profiles yield $\alpha = 1.695\pm0.035$. The exponent for disordered alloys was determined to be $\alpha\simeq 1.667$ in a previous paper \cite{apl-thinfilms}. All three exponents are quite far removed from the ideal limit of Rayleigh scattering. This phenomenon has several causes, from the ever-present influence of three-phonon processes to the fact that, even for alloy scattering, the cross sections only follow the classical $\omega^4$ law at very low frequencies since real phonon densities of states are quite different from the predictions of a Debye model \cite{tamura_isotope_1983}. That the $\alpha$ exponent of Muraki profiles is clearly higher than that of digital profiles and even of the alloy is a more formal statement of the arguments contained in the preceding paragraph.

In quantitative terms, the digital profiles simulated for Fig. \ref{fig:xplane} have between $10\%$ and $20\%$ higher conductivities than the corresponding Muraki profiles. Perhaps more interestingly, they have between $10\%$ and $27\%$ higher conductivities than the random alloys with the same compositions and film thicknesses. This suggests that reducing segregation during growth is a promising path towards higher thermal conductivities, which are interesting for efficient heat dissipation in InAs/GaAs based devices. However, it is experimentally challenging to grow such idealized profiles, since growth would have to proceed at much lower temperatures than typically used, which could result in excess unwanted point defects being present in the sample.
The lower limit for the substrate growth temperature is $460\dg$, since at lower growth temperatures additional As will be incorporated into the structure \cite{Tadayon}. Our superlattices have been grown at $480\dg$, and according to Muraki \cite{Muraki} changes in segregation in the temperature range from there down to $420\dg$ are very limited. Therefore, growth of sharper structures of high-quality MBE InAs/GaAs SL is unlikely. However, low temperature growth with subsequent annealing could be an alternative way to create this kind of structures.

 \section{Conclusions}

 We have theoretically calculated from first principles, and experimentally characterized, the thermal conductivity of InGaAs SLs. The degree of tunability of the bulk thermal conductivity by the structure of the individual periods is much smaller than in the case of SiGe SLs, due to restrictions on how much In can be deposited per period. The total SL thickness, however, strongly impacts $\kappa$, even down to micrometric thicknesses, and the interplay between thickness and composition profile is a major source of tunability for this material. Good agreement is obtained between calculated and measured values for different period lengths, SL lengths, and InAs barrier thicknesses. The measured values confirm the theoretically predicted lack of influence of the segregated nanostructured compositional profiles on $\kappa$. In stark contrast to SiGe based SLs, $\kappa$ of InGaAs SL thin films can in principle also be tuned to be higher than $\kappa$ of the alloy, although this would require being able to grow SLs with virtually no segregation.

\section*{Acknowledgement}

This work has been supported by the European Union's Horizon 2020 Research and Innovation Programme [grant no. 645776 (ALMA)].

\section*{References}

\bibliographystyle{ActaMatnew-2}
\bibliography{bibliography}

\end{document}